\documentclass[conference]{IEEEtran}
\IEEEoverridecommandlockouts
\usepackage{cite}
\usepackage{amsmath,amssymb,amsfonts}
\usepackage{graphicx}
\usepackage{textcomp}
\usepackage{xcolor}

\usepackage{adjustbox}
\usepackage{listings}
\usepackage{booktabs}
\usepackage{multirow}
\usepackage{makecell}
\usepackage{tcolorbox}
\usepackage{multicol}
\usepackage{array}
\usepackage{tabularx}
\usepackage{tabulary}
\usepackage{svg}

\definecolor{codegreen}{rgb}{0,0.6,0}
\definecolor{codegray}{rgb}{0.5,0.5,0.5}
\definecolor{codepurple}{rgb}{0.58,0,0.82}
\definecolor{backcolour}{rgb}{0.95,0.95,0.92}

\lstdefinestyle{mystyle}{
    backgroundcolor=\color{backcolour},   
    commentstyle=\color{codegreen},
    keywordstyle=\color{magenta},
    numberstyle=\tiny\color{codegray},
    stringstyle=\color{codepurple},
    basicstyle=\ttfamily\footnotesize,
    breakatwhitespace=false,         
    breaklines=true,                 
    captionpos=b,                    
    keepspaces=true,                 
    numbers=left,                    
    numbersep=5pt,                  
    showspaces=false,                
    showstringspaces=false,
    showtabs=false,                  
    tabsize=2
}

\def\BibTeX{{\rm B\kern-.05em{\sc i\kern-.025em b}\kern-.08em
    T\kern-.1667em\lower.7ex\hbox{E}\kern-.125emX}}
\begin{document}

\title{An Analysis of Architectural and Operational Dynamics of Phishkits in the Wild}

\author{
\IEEEauthorblockN{Behzad Ousat\textsuperscript{*}, Mohammad Ali Tofighi, Estefan Schafir, Amin Kharraz\textsuperscript{*}\thanks{\textsuperscript{*}Corresponding authors: \texttt{bousat@fiu.edu}, \texttt{mkharraz@fiu.edu}.}}
\IEEEauthorblockA{Florida International University}
}

\maketitle

\thispagestyle{plain}
\pagestyle{plain}

\begin{abstract}
Phishing attacks have always been a favored vector for adversaries to defraud users, bypass modern defense mechanisms, and penetrate critical systems.
Among all the elements contributing to the creation and deployment of successful phishing attacks, phishkits stand out as a crucial parameter. Phishkits often facilitate creating and deploying compelling phishing pages, implement evasion strategies, and establish and maintain backdoors with remote adversaries for exchanging leaked data.
In this work, we performed an analysis of 1,300 modern phishkits collected from 2020 to 2023. We analyzed the architecture, source code, communication channels, and the nature of leaked data shared with adversaries.
We identified mechanisms for dynamic redirection and attributing incoming web traffic as part of the evasion and cloaking mechanism. We also observed heavy reliance on current messaging services for exchanging stolen data with phishers.
That said, our analysis shows that the number of phishkits with advanced functionalities is quite small. We identified 284 (21.8\%) phishkits that did not use any form of evasion mechanism.
We also observed that while there were differences in the implementation details of phishkits, the major components that keep phishing pages functional were very similar or even identical across kits. The level of code reuse and heavy reliance on known tricks to build pre-packaged phishing pages make a large number of cases predictable, which can potentially make the detection of these adversarial operations even easier at scale.
\end{abstract}


\section{Introduction}
Phishing attacks continue to remain a top security threat~\cite{phishkitnews2023-1,phishkitnews2023-2}. These attacks have evolved significantly in scale and sophistication and their impacts have become deeper and more consequential.
One question that arises is how adversaries maintain their asymmetric power to generate phishing websites at scale. Part of the reason is that the economy of mechanisms~\cite{phishkitnews2023-3,phishkitnews2023-4,phishkitnews2023-5} in developing phishing websites has improved significantly over the years.
In particular, to develop effective phishing websites, adversarial campaigns often use pre-packaged software tools and templates, referred to as phishkits, to develop, maintain, and manage phishing websites.
Modern phishkits are optimized in several ways. With a usable command line interface, adversaries can generate phishing websites that satisfy the look and feel of the page and implement fingerprinting~\cite{shirazi2020machine,roy2022survey,shirazi2023adversarial}, evasion~\cite{liu2023knowledge,apruzzese2022spacephish}, and cloaking~\cite{li2023uncovering,zhang2021crawlphish,fujii2023stargazer,liu2023knowledge,apruzzese2022spacephish,zhang2022m} best practices.
The ease of use and accessibility of phishkits have significantly reduced the cost of launching large-scale phishing campaigns, allowing even those with limited technical expertise to deceive unsuspecting victims.

In this paper, we aim to explore how phishkits are used to generate real-world social engineering attacks, how the core components of phishkits reduce the cost of developing phishing websites, and how effective those approaches are. We also investigate what forms of evasion techniques are being used. Answering these questions can potentially help us, as defenders, adapt our defense mechanisms for more robust detection. To perform the analysis, we created a dataset of 1,300 phishkits seen between 2020 and 2023. We collected 589 phishkits by searching public repositories~\cite{OpenPhishCom,phishhuntio,DLphishkitsgithub} and 732 unique phishkits obtained through a partnership with a well-known web security company that gave us daily access to the source code of phishing websites.
In the following, we highlight some of the major results of this paper:

\textbf{First, our analysis shows that the core functionalities of many phishkits are predictable and follow specific engineering practices}. That is, to build effective phishing websites, modern phishkits should not only satisfy the look and feel requirements to generate successful phishing pages but also implement evasion mechanisms (e.g., traffic attribution and management) and backdoor management. Using the data, we identified several instances for each of those mechanisms. For instance, we identified 452,717 unique blocked IP addresses/ranges that belonged to local ISPs or known security companies, search engines, specific universities and institutions, and VPN providers -- most likely as a mechanism to defend against web scanning and crawling operations.
We also identified 353 unique Telegram bot tokens and 456 email accounts that were used to share the leaked data with adversaries.

\textbf{Second, secure messaging APIs are extensively being abused by the majority of phishkits to transfer extracted information automatically}. We interacted with 292 Telegram bots we extracted from the phishkits by sending periodic poll update requests to the bots to verify if they were still active and if they responded to our requests.
We received 8,635 Telegram messages from Telegram bots over the experiment period. We identified 370 leaked credit card information, 425 SMS verification codes for 2-Factor Authentication, and 509 username/password pairs in known financial institutions and top 500 Fortune companies. We responsibly shared the collected information with the anti-fraud department of major credit card issuers seven months before submitting the paper.
We received acknowledgment that all the active cards had been canceled and users were notified by the card issuers.

\textbf{Third, our analysis shows that most of the identified phishkits follow previously known and straightforward techniques to build their phishing pages.} For instance, we observed that
126 phishkits (9.6\%) used identical IP blocklists with almost no changes in the content.
We also observed that the evasion mechanisms by and large did not have any critical changes over the years. In line with that observation, we also identified 284 phishkits that did not contain any mechanism for evasion purposes, providing access to any request, making detection much easier by current solutions such as Google Safe Browsing~\cite{GoogleCPW}.
Considering other aspects of the analysis in evasion, we conclude that the majority of phishkits we observed in this dataset were following simple mechanisms for deployment and their core functionalities to deliver phishing pages were highly predictable -- making their detection even easier.

Given the number and type of leaked information in the experiment, our hope is that
this work serves to raise awareness about phishkits and the importance
of defining systematic approaches to improve defense agility against these large-scale threats that are routinely
observed in the modern web.
We hope that the insight in this paper can potentially open new avenues on the defense side to develop automated tools to effectively model the predictable behaviors in phishkits
to locate them with minimal human intervention in the loop at scale.

\noindent \textbf{Contributions.} In summary, the contributions of the paper are as follows:
\begin{itemize}
    \item We analyzed 1,300 phishkits, describing less well-studied aspects of phishkit platforms in the wild with a focus on evasion mechanisms and common characteristics among different phishkits.

    \item We explain how the core components of phishkits are engineered and how they are being used in the entire dataset.

     \item We investigate the communication channels across pre-packaged phishing websites and perform
     an analysis of the leaked data.


\end{itemize}

\section{Background}
\label{sec:background}

\subsection{General Architecture of a PhishKit}
Phishkits often follow a set of practices to stay effective. In the following, we have categorized these functionalities based on the analysis of the collected samples (discussed in Section~\ref{subsec:Longitudinal Dataset}).

\noindent \textbf{App Template Loader.} App templates are pre-packaged web pages designed to mimic legitimate websites, such as bank login pages, email portals, or social media login screens. Phishkits often have several templates already included in the package, enabling adversaries to deploy different front-ends with minimal effort.
The template loader module is also responsible for generating different versions of the same website. That is, as target websites update their interface, add new services, modify visual properties, or implement new features, the app loader module also adapts the phishing page in order to ensure that the phishing page looks current and convincing.

\noindent \textbf{Backdoor Management:} Phishkits often have integrated communication channels to share the leaked information with adversaries.
To maximize effectiveness, the underlying technology should offer low-cost evasion while satisfying the confidentiality requirements of the channel so that the exchanged data
may not be disclosed to third parties. To this end, phishing campaigns often use or establish mail server infrastructures that are served over TLS to exchange data over the SMTP protocol.
Modern messaging technologies have become a robust alternative, as generating new messaging bots is a fairly straightforward operation, carried over secure channels, and can easily bypass some of the contemporary blocklisting mechanisms.

\noindent \textbf{Traffic Attribution:} A successful phishing website requires more than what we refer to as the look and feel property to stay effective. For instance,
the web server attributes the incoming traffic before delivering the actual phishing page.
Phishkits often have modules to facilitate this operation by employing various tactics to reason about the incoming traffic. The module often compares the origin network address to infer any links to search engines, security companies, government entities, or universities actively involved in web scanning.
Furthermore, they employ multifaceted client validation methods, including rDNS queries and user agent checks, often employing a combination of these techniques.

\begin{figure*}[h!]
    \centering
    \includegraphics[width=0.7\linewidth]{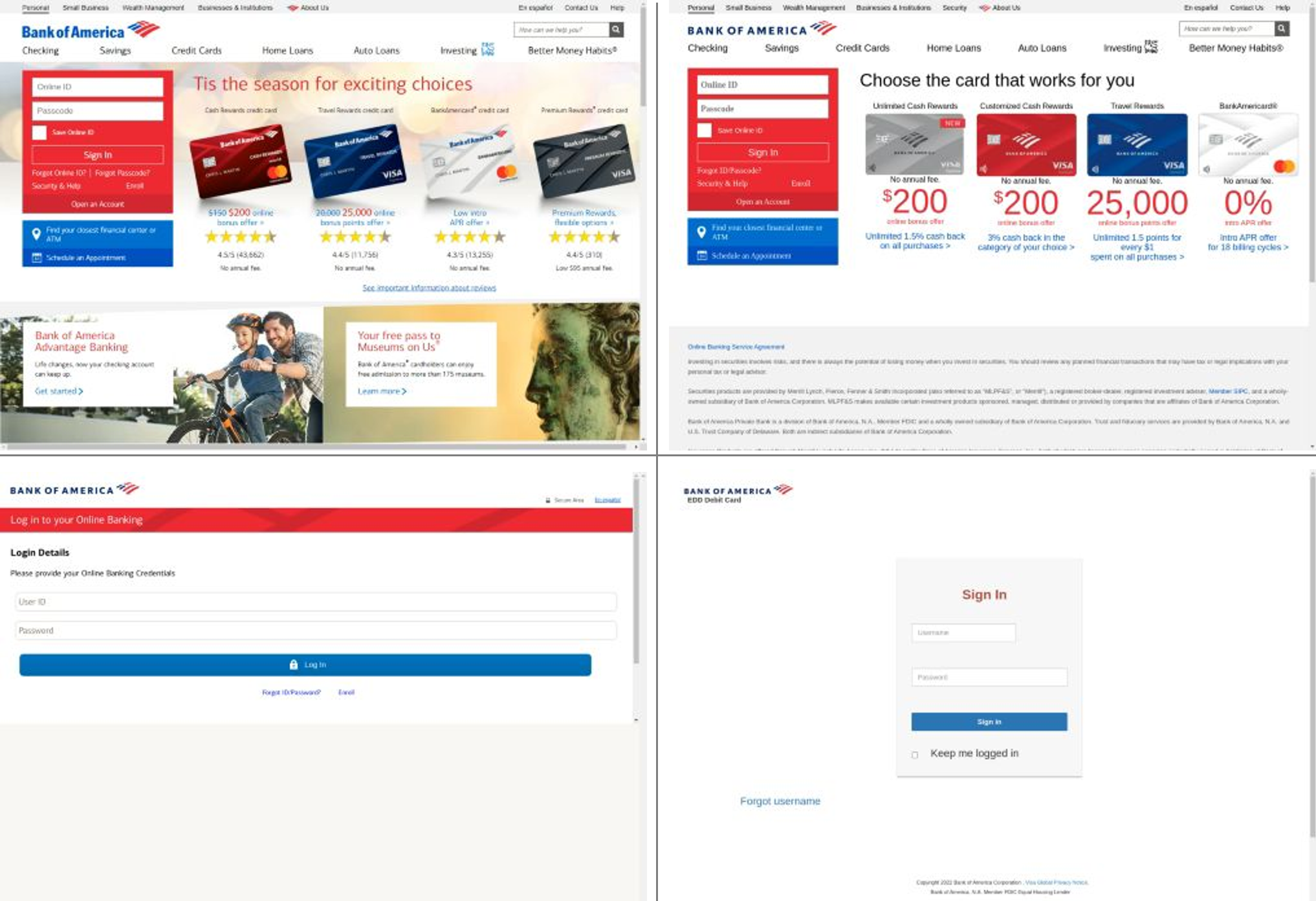}
    \caption{Fraudulent website created using a phishkit. These fake websites are clones of the legitimate entity and contain a login form. If users input their credentials, the data will be sent to the attacker.}
    \label{fig:motivating_example}
\end{figure*}

\noindent \textbf{A Motivating Example: }
In this section, we provide a real-world example where a high-profile website, i.e., Bank of America, is being targeted by several phishkits we observed in our dataset.
We identified 22 phishkits that were used to generate fraudulent versions of the Bank of America login or main page where users had to enter their username and password.
Each phishkit uses a multitude of different subnets and hosting providers. We identified 8 distinct ASNs that correspond to 12 IP addresses.
In addition to using relatively diverse infrastructure, the domains were hosted over CDNs, rendering
blocklisting techniques less effective. We observed three Telegram bots and 10 email addresses that were used to share the leaked information with the attackers.
Our observation revealed that half of these instances involve the use of inexpensive top-level domains (TLDs), such as .vn, .info, and .online, to generate
different variations of addresses for their domains as well as email servers.

\subsection{Prior Research on Phishkits}

Phishkits are not a new concept.
One of the first studies of phishkits was conducted by Cova et al.~\cite{cova2008there} in 2008, where they performed an analysis of the backdoor mechanism in phishkits to establish remote communication with adversaries.
Han et al.~\cite{han2016phisheye} performed an analysis of live phishing websites created using phishkits in a honeypot environment.
Their analysis included 643 phishkits that were used by more than 470 attackers. They showed that phishkits were short-lived and had been active for less than 10 days, and that block-listing mechanisms were not very reliable in minimizing the impact of this form of malicious code. More recently, Bijmans et al.~\cite{274537} performed an analysis of 70 Dutch phishkits and evaluated their prevalence in the wild. The authors performed an end-to-end life cycle analysis -- from domain registration and
kit deployment to take-down.
Bijmans et al.~\cite{bijmans2021catching} analyzed information-sharing flows throughout phishing websites, including how information from phishing sites is transmitted to
attackers and third parties. However, they only considered email as the transmission technology.
Peng et al.~\cite{10.1145/3321705.3329818} introduced a fingerprinting scheme that they used to cluster similar kits together, using a dataset of 2,064 phishkits. They used uncommon file names to fingerprint each kit and formed different families of phishing kits, sometimes connected to each other. Those connected families can show evolution and code reuse among phishkits.

\noindent \textbf{This Work.}
This paper is the first work to conduct a longitudinal study of the development details of phishkit samples from 2020 to 2023. The analysis aims to answer what features have changed and how code samples have evolved over time. We also analyze what features are being reused among different phishkits and how those artifacts can be used to detect phishing pages developed by the same phishkits.

\section{Methodology}

\label{subsec:Longitudinal Dataset}

\noindent \textbf{Data Collection Method:}
One approach to collecting phishkits relies upon deploying deception-based infrastructure, allowing adversaries to compromise internet-exposed machines and use them to host their phishing infrastructure. This allows defenders to collect real-world evidence about phishkits and other forms of attacks.
We partnered with a well-known web security company and obtained 732 unique phishkits from its daily feed; the collection period for this feed extended from December 21, 2022, to March 20, 2023. We additionally collected 589 phishkits from the Phishunt public repository~\cite{DLphishkitsgithub}. After removing duplicate and irrelevant samples, and accounting for 21 kits shared between the two sources, the final dataset contained 1,300 unique phishkits observed between 2020 and 2023.

\noindent \textbf{Generalizability of the Collected Dataset:} We considered three orthogonal metrics to reduce potential biases. First, the phishing kits in the dataset do not solely focus on English websites and cover 23 different languages. Second, the phishing kits used in the experiment have been observed over the 2020--2023 observation period, covering possible evolution and temporal changes. Third, the collected phishing kits target several top enterprises, such as financial institutions (e.g., Bank of America, Wells Fargo, PayPal), governments, airlines, postal services, the healthcare industry, and insurance and loan companies. Lastly, the dataset includes items from both a premium data feed and a publicly available phishkit dataset to better mirror what exists in the wild. While we considered these metrics, we still believe that our analysis could be impacted by other forms of bias (code reuse among adversaries, not enough cases from all languages or geographical locations, or undetected kits that may be more sophisticated). We should note that the process has been a best-effort approach, given all the limitations to acquiring real-world malicious datasets. To offer a comprehensive depiction of the entities targeted by phishing attacks, we have categorized them based on their primary activities or sectors of operation. The data for each category is presented in Table \ref{tab:data_collection_summary} for clarity and reference.
The categories were defined manually after consulting publicly available services (e.g., Similarweb) as well as the names of the institutions.

\begin{table}
\centering
\caption{Most commonly observed business sectors that are targeted by phishkits.}
\label{tab:data_collection_summary}
\scriptsize
\begin{tabular}{l r r}
\toprule
\textbf{Category} & \textbf{\# Phishkits} & \textbf{\# Domains} \\
\midrule
Generic/Spear Phishing& 255& 351
\\
Bank& 186& 205
\\
Cloud Services& 151& 213
\\
Credit Unions& 112& 147
\\
Social Network& 88& 97
\\
Postal Services/Package Delivery& 77& 76
\\
Streaming Services& 67& 67
\\
Industry& 55& 54
\\
Shop& 40& 41
\\
Mail Services& 40& 48
\\
Government& 28& 62
\\
Telecom& 23& 42
\\
Online Services& 7& 7
\\
Medical Services& 3& 3
\\
Other& 168& 164\\
\midrule
\textbf{Total} & \textbf{1,300}& \textbf{1,577}\\
\bottomrule

\end{tabular}

\end{table}

\noindent \textbf{Ethics.}
Part of our analysis was to interact with Telegram bots that were hardcoded in some of the phishkits. One of the research questions was to determine if those Telegram bots would respond to requests and, if they did, what kind of information could be retrieved. While no human subjects were involved in the process, working with those bots could raise some ethical concerns because they would leak information about victims, including PII such as email addresses, names, and credit card information.

We created a case for review with the Institutional Review Board (IRB) office for this experiment and shared the entire experiment pipeline.
We received approval to interact with the bots to complete this study. We were also advised not to conduct any research on any of the leaked PII.
\textbf{We treated all the retrieved data (e.g., IP addresses, emails, card information, physical addresses) as PII. We did not use the identifiable information in any form in any part of this research.}
Data protection procedures have been implemented
at all layers. The collected data was aggregated and encrypted via a public key on a local machine. The machine is only accessible to two members of the team who
had successfully passed special IRB training courses to work with PII data.
To minimize any future attempts to interact with those Telegram bots, we intentionally removed all the Telegram bots from the dataset we made available.
Also, we removed a set of log files that contained the IP addresses of connected clients in the collected phishkits.
\textbf{Lastly, we contacted the Online Payment Fraud department of a major credit card issuing company seven months before releasing the paper.} We shared the collected dataset over their secure file transfer service for them to begin the investigation and damage control process.
We also contacted the Telegram abuse department several times over the last seven months to discuss the findings and share the offending Telegram bots; however, we did not receive any response at the time of submission.

\section{Architectural and Operational Dynamics in Phishkits}
In this section, we study the implementation of the three major components of phishkits in the dataset we collected. We aim to identify common practices in terms of sophistication across different phishkits, as well as possible similarities across the samples in terms of evasion and attribution.

\noindent \textbf{Backdoor Mechanisms:}
As mentioned earlier in Section~\ref{sec:background}, communication with the remote phisher occurs in different forms.
One classic approach is the use of email services to automatically craft an email containing leaked information to share with phishers. While we identified 456 public or private mail services across all the phishkits, we observed that Gmail has been the most abused mailing service by phishers, with a wide margin over the rest of the dataset. That is, we identified 279 unique Gmail addresses in 299 phishkits. We also identified 20 email addresses that were being used by more than one phishkit. Other popular mailing services were Yandex (66), Protonmail (23), Yahoo (22), and Hotmail (13).
We also observed the use of Telegram bots~\cite{TGLBotRev} as an alternative method to exchange leaked information with phishers through standard APIs.
There are several reasons that make this service an attractive target for adversaries. One is the anonymity feature, which makes it difficult for regulatory bodies and legal entities to trace the creator of the bot. The more important reason is the low cost of evasion, because generating hundreds of bots is a fairly simple process.
In particular, generating new Telegram IDs is programmable and can be done automatically, introducing almost zero operational cost.
By applying regular expressions on the source code of the phishkits, we identified 292 Telegram bot tokens in 314 (42.9\%) of the 732 live kits from our daily feed, and 61 tokens in 60 (10\%) of the 589 kits available in the Phishunt dataset. We also identified one phishkit that was set up to use a Discord channel to send the stolen data. However, we did not find a strong correlation between this observation and other parts of the phishkit in terms of utilizing more advanced techniques.

\noindent\textbf{Automatic Interaction:}
We had close to real-time access to the premium dataset feed. That is, the list of newly found phishing kits was updated every five minutes. Consequently, the chance that a given Telegram bot had been active was relatively high.
We used Telegram's \texttt{getUpdates} method to poll updates from those bots and process the information they might have received. We queried the bots every 3 hours during a 101-day period and captured a total of 8,635 messages. The output of the tokens contained a significant amount of sensitive user input, encompassing username/password combinations, SMS verification codes for 2-Factor Authentication, and credit card details.
In particular, we identified 370 credit cards across 33 unique phishkits, and among them, 243 (65.7\%) were issued by one entity.

\begin{samepage}
\noindent\textbf{As mentioned in the Ethics section, we did not perform any analysis of the leaked data, nor can we disclose more information on the leaked payloads.}
\end{samepage}

\section{Traffic Attribution}

Our analysis shows that cloaking mechanisms~\cite{cloakx-sec19,deng2013uncovering,duan2017cloaker} are a common practice in phishkits.
Cloaking allows adversaries to avoid being easily spotted by automated scanners and search engines that are designed to crawl the web to identify potential abuses or update blocklists (e.g., Google Safe Browsing) to reduce user exposure to these threats. In this section, we investigate common practices that phishkits use to attribute incoming traffic and perform cloaking or redirection in
response to requests originating from specific IP addresses or user agents.
In the course of the experiment, we observed that 600 (46.1\%) of the phishkits were incorporating common IP blocklisting techniques to filter unwanted traffic. This involved blocking specific IP addresses or IP ranges associated with search engines, government agencies, security companies, research institutions, and specialized anti-phishing services. Listing~\ref{fig:bofa-evasion} shows a real sample, observed among the phishkits targeting Bank of America, of a phishkit that directly checks the client's IP address against a static list of banned addresses before serving the phishing page.

\noindent\begin{minipage}{\linewidth}
\begin{lstlisting}[language=PHP, caption={An evasion technique used in phishkits to target Bank of America}, label={fig:bofa-evasion}]
<?php
$hostname = gethostbyaddr($_SERVER['REMOTE_ADDR']);
$bannedIP = array(
    "<List of ip addresses>"
);
if(in_array($_SERVER['REMOTE_ADDR'],$bannedIP)) {
    header('HTTP/1.0 404 Not Found');
    exit();
} else {
    foreach($bannedIP as $ip) {
       if(preg_match('/' . $ip . '/',$_SERVER['REMOTE_ADDR'])){
         header('HTTP/1.0 404 Not Found');
           die("<h1>404 Not Found</h1>
    The page that you have requested could not be found.");
        }
     }
}
?>
\end{lstlisting}
\end{minipage}

Apart from directly blocking IP addresses, 312 (24\%) used the client's IP address to look up their ISP and blocked specific ISPs, or any ISP whose name contained specific words. Listing~\ref{fig:isp-sample} shows a real sample of a phishkit that blocks any client with the word ``University'' in the ISP owner's organization name. This way, they do not need to maintain a list of blocked IP addresses and can dynamically block new IPs from universities as they reach their web page. This specific phishkit uses a free tool to get meta information from an IP address.

\noindent\begin{minipage}{\linewidth}
\begin{lstlisting}[language=PHP, caption={Blocking all university IP addresses},label={fig:isp-sample}, basicstyle=\scriptsize]
function whois(){
return @json_decode(file_get_contents("http://ip-api.com/json/" . getIp()), true);
}
$badISPs = array(
    ...
    "University",
    ...);
$isp = $whois()["isp"];
foreach ($badISPs as $badISP) {
    if (preg_match("/" . $badISP . "/i", $isp)) {
        logger($badISP, "BOT ISP");
        die(header("Location: https://www.google.com"));
    }}

\end{lstlisting}
\end{minipage}

Table \ref{tab:kits_ip_blocking} shows a summary of IP blocking by phishkits.
We identified traces of IP blocklisting in 600 of the collected phishkits. We also observed that 284 phishkits did not have any form of evasion mechanism in place, leaving their phishing web page open to scanning and analysis, and that 439 phishkits were incorporating some level of attribution and fingerprinting based on user agent analysis. Listing \ref{fig:phishkit-evasions} illustrates a code sample showing how the PHP code analyzes the user agent and IP address of the incoming traffic to redirect unwanted clients. We should note that the technique used for evasion is fairly well known and is not necessarily used only by phishkits. Other forms of scams have been reported to use these tactics in prior work~\cite{kharraz2018surveylance,MiramirkhaniSN16}.

\begin{table}
\centering
\caption{Summary of analyzing IP Blocking in phishkits}
\label{tab:kits_ip_blocking}
\begin{tabular}{l r r}

\toprule
\textbf{Phishkits} & 1,300 \\
 \midrule
 \textbf{Phishkits with IP Blocking} & 600\\
 \midrule
 \textbf{Total IP Address/Ranges Blocked} & 5,092,303 \\
 \midrule
 \textbf{Unique IP Ranges Blocked} & 452,717\\
 \midrule
 \textbf{Unique IP Addresses Blocked} & 358,561\\
\bottomrule

\end{tabular}

\end{table}

\noindent\begin{minipage}{\linewidth}
\begin{lstlisting}[language=PHP, caption={Evading crawlers from search engines, suspicious IP addresses, and phishing detection services}, label=fig:phishkit-evasions]
$bots_agents = array('googlebot', 'yahoo', 'yandexbot');
$bots_ips = array('1.2.3.4', '2.3.4.5');
$blocked_words = array(
  "phishtank", "google", "tor-exit", "amazonaws");
$user_agent = $_SERVER['HTTP_USER_AGENT'];
$user_ip = $_SERVER['REMOTE_ADDR'];
$hostname = gethostbyaddr($_SERVER['REMOTE_ADDR']);
foreach($bots_agents as $bot_agent) {
  if (substr_count($user_agent, $bot_agent) > 0)
    exit(header('Location: https://www.masrawy.com'));}
foreach($bots_ips as $bot_ip) {
  if (preg_match('/'.$bot_ip. '/', $user_ip))
    exit(header('Location: https://www.masrawy.com'));}
foreach($blocked_words as $word) {
  if (substr_count($hostname, $word) > 0)
    exit(header('Location: https://www.masrawy.com'));}
\end{lstlisting}
\end{minipage}

\begin{table}
\scriptsize
\centering
\caption{Distribution of blocked organizations by phishkits. Cloud and security providers are among the most commonly blocked organizations across the observed phishkits.}
\label{tab:top-blocked-orgs}
\resizebox{\linewidth}{!}{%
\begin{tabular}{l l r r}

\toprule
\textbf{Type} & \textbf{Organization} & \textbf{\# Kits} & \textbf{\# IP Addresses} \\
\midrule
\multirow{8}{*}{\makecell{ISP, Cloud \\ \& Security Providers}}
 & Tri-County Communications Cooperative, INC.               & 440 &                         2 \\
 & Google LLC                                                & 412 &                      7,492 \\
 & TalkTalk Communications Limited                           & 348 &                       158 \\
 & Microsoft Corporation                                     & 347 &                      6,791 \\
 & Hetzner                                                   & 325 &                      4,197 \\
 & AWS EC2 (us-east-1)                                       & 314 &                     25,634 \\
 & AWS EC2 (eu-west-1)                                       & 311 &                     15,559 \\
\midrule
 \multirow{13}{*}{Research Institutions}
 & Universiteit van Tilburg                                     & 178 &                         3 \\
 & University of Iowa                                           & 172 &                         3 \\
 & City University of Hong Kong                                 & 169 &                         4 \\
 & University of Idaho                                          & 168 &                        30 \\
 & University of California, San Diego                          & 143 &                        14 \\
 & Technische Universiteit Eindhoven                            & 133 &                        10 \\
 & Norwegian University of Science and Technology               & 132 &                        27 \\
 & Massachusetts Institute of Technology                        & 131 &                        13 \\
 & Carnegie Mellon University                                   & 130 &                         3 \\
 & University of Washington                                     & 129 &                        10 \\
 & Universitaet des Saarlandes                                  & 128 &                        14 \\
 & Stanford University                                          & 128 &                         5 \\
 & Computer Center of Peking University                         & 128 &                         3 \\
 \midrule
\multirow{7}{*}{VPN Services}
 & BullGuard ApS                                              &  235                        & 2 \\
 & VPN Consumer Network                                        &  33                       & 10 \\
 & NordVPN                                                     &  31                        & 4 \\
 & Private Internet Access, Inc                                &  29                       & 19 \\
 & ForPrivacyNET                                               &  26                        & 7 \\
 & The PRIVACYFIRST Project                                    &  19                        & 1 \\
 & Anonymouse                                                  &  18                        & 4 \\
\bottomrule

\end{tabular}
}
\end{table}
We performed an analysis of over 358,561 unique IP addresses we found in all the collected kits to identify the organizations or entities those phishkits tend to avoid. A summary of blocked organizations is shown in Table~\ref{tab:top-blocked-orgs}. The majority of blocked IPs belong to cloud and security companies. A more interesting finding was that a large portion of phishkits were using very similar, and in some cases exactly the same, evasion packages. For instance, we observed that 126 of the phishkits were using a set of identical files with identical names to evade unwanted traffic, all imported together at the same time.
We also identified a large list of research institutions that were blocked by the phishkits. Table~\ref{tab:top-blocked-orgs} provides the top institutions we observed in the lists.
We identified several research institutions that are very active in web measurements and security scanning, whose IP addresses were listed in a majority of the phishing kits.
One question that arises is why VPN services are blocked by phishkits.
While answering this requires more data points to make any scientific claim, our analysis of the cases suggests that blocking VPNs could happen for at least two main reasons: (1) it is common practice for security researchers to use VPN services in their crawling experiments to disguise their real IP addresses; by filtering out VPN traffic, phishkit operators attempt to prevent these attempts from analyzing and detecting their phishing operations, and (2) phishkits may be designed to target users in specific countries or regions -- filtering VPNs helps ensure that the traffic is genuinely from the targeted locale, as VPNs can mask a user's actual geographic location.
We also observed other uses of third-party services for traffic attribution. We identified the use of geolocation services such as \texttt{geoplugin.net} and \texttt{ip-api.com} to deliver content to specific geographical locations. We observed that 498 phishkits were using the geoplugin service and 120 phishkits were using ip-api to fetch the location of the connecting agent.
Listing~\ref{fig:phishkit-geo-evasion} shows how the phishkit makes a request to the geoplugin service and decides whether to respond to the HTTP request with the actual phishing page.

\noindent\begin{minipage}{\linewidth}
\begin{lstlisting}[language=PHP, caption={A sample code used to allow only users from South Africa to a fraudulent website}, label=fig:phishkit-geo-evasion]
$ip = $_SERVER['REMOTE_ADDR'];
$details = json_decode(file_get_contents("http://www.geoplugin.net/json.gp?ip=$ip"));
$country = $details->geoplugin_countryName;
if ($country !== "South Africa") {
  header("Location: https://online.banking.com/login");
  exit();
}
\end{lstlisting}
\end{minipage}

Along with the typical evasion methods discussed above, we identified 140 (10.7\%) phishkits that were implementing a form of moving target defense to prevent URLs from being detected and blocklisted by popular systems like Google Safe Browsing~\cite{GoogleSB}. We observed that they make a copy of their entire source code into a randomly generated path and redirect users to that path. This approach would clearly fail if the root URL gets detected; however,
it could potentially work well in an environment where blocking the domain name would introduce significant cost. A sample of the snippet used to copy the entire website files to a random path is shown in Listing~\ref{fig:phishkit-recurse-copy}.

\noindent\begin{minipage}{\linewidth}
\begin{lstlisting}[language=PHP, caption=A sample code used to copy the entire files of a fraudulent website to a random path, label=fig:phishkit-recurse-copy]
function recurse_copy($src, $dst) {
    $dir = opendir($src);
    @mkdir($dst);
    while (false !== ($file = readdir($dir))) {
        if ($file != '.' && $file != '..') {
            if (is_dir($src . '/' . $file))
                recurse_copy($src . '/' . $file, $dst . '/' . $file);
            else
                copy($src . '/' . $file, $dst . '/' . $file);     }   }
    closedir($dir);}
$src = "app";
$random = rand(0, 9000000);
$md5 = md5($random);
$base = base64_encode($md5);
$dst = strtolower(md5($base));
recurse_copy($src, $dst);
header("Location: $dst");
\end{lstlisting}
\end{minipage}

\section{Analysis of Phishkits Over the Years}
To investigate the changes over the years,
we performed an analysis of the source code of extracted phishkits.
While a phishkit can contain several files, including CSS and JS, we primarily focused on the PHP source code in the phishkits, as it serves as the core function in generating phishing websites. We generated the control flow graph (CFG) of all the collected PHP source code by generating the abstract syntax tree (AST) for each sample and extracting the corresponding control flow information.
For each phishing kit, we converted the call graph and CFG into feature vectors and constructed an adjacency matrix that described the relationships (calls, control flow) between functions and blocks in the code.
Since adjacency matrices are high-dimensional and may not directly capture structural semantics, an alternative is to embed them into a lower-dimensional space before computing distances.
After evaluating various approaches, including spectral embedding~\cite{DBLP:journals/corr/abs-2009-14441}, Node2Vec~\cite{DBLP:journals/corr/GroverL16}, and graph neural networks (GNNs)~\cite{DBLP:journals/corr/abs-1812-08434}, we selected Node2Vec for converting adjacency information into node embeddings. This choice was driven by its ability to operate in an unsupervised manner and its computational efficiency, as opposed to GNNs, which require iterative message passing across the graph.
We computed pairwise cosine distance on the embeddings and applied agglomerative clustering~\cite{Zepeda-Mendoza2013} to generate clusters of similar phishkits.

We identified 171 similar clusters, with 68 clusters having at least 3 phishkits.
Although phishkits had differences in their CSS, JS, and HTML code, the generated clusters show that 276 kits (21\%) had structurally identical layouts and 100\% structural similarity.
To get a flavor of the similarities among phishkits in clusters, we took a closer look at the largest cluster, which contained 17 phishkits.
The oldest phishkit in this cluster was discovered in November 2022, while the most recent one was detected in late February 2023. These kits were targeting banking systems, including Wells Fargo and Bank of America, and were highly similar, with similarity scores ranging from 82\%--100\%.
All of the phishkits were using the same code sample to interact with the remote mail servers when sharing the leaked information.
However, upon checking the contents of these kits, we noticed small differences in the messages shown to victims, and that some kits had added or removed intermediate pages. These minor differences can fool detection systems that rely on the visual similarity of detected phishing websites, or on the behavior of the client when interacting with them. Such modifications can be made intentionally for these purposes, or can arise unintentionally over time as kits are kept similar to the original website. The differences also included destination addresses for stolen data.

\section{Discussion}
In this
section, we discuss the implications of our investigation for the state of the art. We describe how the findings in this paper can inform other activities on the defense side, and how the security community should
react moving forward to new emerging threats of this sort.

\noindent \textbf{Enhancing Auditing Processes}
Our analysis unveiled important trends in the abuse of current technologies
among phishing campaigns. For instance, we observed the adoption of low-cost evasion techniques
to enable backdoors to adversaries' machines, similar to the techniques used in malware attacks~\cite{lever2017lustrum}, making blocklisting more difficult. Furthermore, our analysis confirms the folk wisdom that secure messaging platforms are widely abused by adversaries to share stolen credentials. This trend calls for more coordinated efforts to define rigorous auditing mechanisms for the use of legitimate services for adversarial purposes. We acknowledge that establishing such coordination has never been easy.
However, cloud service providers, credit card issuers, domain registry operators, and browser vendors need to reach
consensus on
where and how auditing mechanisms should take place at web scale.
While cloud service providers have become increasingly efficient at preventing phishing pages on their infrastructure, the main line of defense protecting users against web-based social engineering attacks is still browser vendors. We believe stronger coordination between these entities is critical to responding more quickly to these attacks and protecting Internet users.

\noindent \textbf{Improving Zero-Day Identification}
Phishkits are indeed an important part of phishing attacks, allowing
phishing campaigns to launch their attacks more cheaply and quickly. However, our insights from the analysis show that using phishkits to develop fraudulent websites can also lead to easier detection. In fact, the heavy reliance on phishkits can have side effects on the adversaries' side by making the behavior and structure of phishing websites significantly more predictable.
Consequently, unsupervised or semi-supervised techniques that can reason about the similarity of website structure at scale can bring significant visibility into these practices. We are aware that major cloud services are moving towards more effective anti-fraud solutions, and detecting this form of phishkit-based social engineering is not likely to remain very difficult. However, there do exist fundamental gaps in detecting
temporal drifts and ``patient zero'' phishkits -- those that are out-of-distribution and have not been seen in prior observations.
This requires further systemic investigation into how to frame patient-zero threat detection as an unsupervised or semi-supervised problem, making it
more likely to identify trends at a global scale.
Development of these techniques potentially allows defenders to build a valuable
knowledge base of modern phishkits and their behavior over time, which
can be helpful in several ways, such as detecting new adversarial practices, identifying threat campaigns, and performing longitudinal measurements on the dynamics of this arms race.

\section{Limitations}

\noindent \textbf{Generalizability of the Insights.} Our analysis pipeline incorporates multiple vantage points to look
at the phishkit ecosystem. It is critical to discuss limitations and
specific systematic biases that might have had an impact on the
robustness of the method and the generalizability of the findings.
We discuss some potential risks in running
experiments similar to ours.
Our approach for creating clusters was based on the network address and the company names.
We made our judgment based on the results of reverse DNS lookups.
While there are public and premium services (e.g., Similarweb~\cite{SimWeb}) that provide the exact industry for a given IP address, we observed that the records in those services do not include many cases, including local ISPs and VPN services. Consequently, our analysis of target sectors or blocked institutions is based on almost 50\% of the collected network addresses. Furthermore, our approach to collecting
phishkit samples is also a best-effort approach. We incorporated 1,300 phishkit samples from public repositories as well as a premium service to get a balanced combination of what is available in
a study covering samples observed from 2020 through 2023.
However, we cannot claim this dataset represents all the cases seen in the wild, and our conclusions are drawn solely from the observations we have from this dataset and should be viewed as a lower-bound
estimate of the full phishkit ecosystem.

\noindent \textbf{Phishkit-based Websites vs. Other Social Engineering Attacks.} The scope of this paper is limited to the common practices seen in phishkits used to launch
phishing attacks.
We do not have sufficient data and evidence to measure how effective these attacks are at exfiltrating user data compared to other forms of social engineering attacks. Another limitation of our study is that we cannot provide thorough information on how the leaked data is used by adversaries. While phishing campaigns, like other scammers, have
significant freedom in how they use the collected user information, we
cannot provide accurate insights into how they may utilize
the collected data. We have not investigated the interactions
of involved parties to make statistically significant claims about
particular types of personal information misuse.

\section{Related Work}

\noindent \textbf{Deceptive Practices in Social Engineering Attacks.}
Prior work has studied evasion mechanisms in modern scams and social engineering attacks~\cite{salahdine2019social,ABRAHAM2010183,266523,2810103.2813724,2897845.2897918,covid-themed,zhang2021crawlphish,tofighi2024constructs}. In particular, Invernizzi et al.~\cite{45365} studied different browser-based and network-based cloaking techniques used to allow only desired traffic. Oest et al.~\cite{8376206} used server-side cloaking techniques as a feature to fingerprint phishkits. In another study~\cite{8835369}, the authors deployed artificial phishing websites equipped with cloaking techniques to evaluate the effectiveness of state-of-the-art detection systems. Zhang et al.~\cite{zhang2021crawlphish} also specifically analyzed client-side cloaking techniques used in phishkits. Miao et al.~\cite{10.1145/3576915.3623199} discussed the vulnerability of Google Safe Browsing and generated evasion mechanisms to evade its visual phishing detection.
While some of the analysis in this paper is in line with prior work (e.g., fingerprinting~\cite{Nikiforakis,Eckersley,10.1145-3,10.1145-1}, cloaking~\cite{duan2017cloaker,deng2013uncovering,invernizzi2016cloak}), this paper primarily focuses on the common characteristics of phishkits, as well as the ecosystem in which phishkits operate to achieve adversaries' goals.

\noindent \textbf{Studying Social Engineering Attacks.} There is a large body of work studying social engineering and its deceptive practices in the wild. Prior work has discussed the underlying ecosystem of different web-based social engineering attacks, such as survey scams~\cite{kharraz2018surveylance}, technical support scams~\cite{liu2023understanding,nirwan2023comprehensive,adu2022phishing,dam2019large}, and fraudulent e-commerce websites~\cite{bitaab2023beyond}.
In this work, we primarily focus on the tools adversaries use to implement successful social engineering attacks faster and at scale. We analyzed these tools considering the sophistication and complexity of the techniques and the potential consequences for defense.

\noindent \textbf{Development of Phishing Websites.} Prior research has focused on several aspects of the development of phishing websites. Hao et al.~\cite{hao_2013} and Garera et al.~\cite{garera_2007} focused on the domain registrars and URL patterns used in the deployment of phishing websites. Additionally, studies such as~\cite{covid-themed, kharraz2017techniques, 10754708, hoheisel_2022} provide insight into how social disturbances can motivate adversaries to develop novel phishing websites. The work most relevant to ours is Bijmans et al.~\cite{274537}, which focuses on the lifecycle of phishkits in the wild. Compared to this work, we performed a more comprehensive study of phishkits over three years, aiming to provide a more holistic picture of the tactics, techniques, and procedures (TTPs), and evidence of the choke points and predictable behaviors that could make detection easier.

\vspace{-0.4cm}
\section{Conclusion}

In this paper, we conducted an analysis of 1,300 phishkits observed from 2020 to 2023. We particularly examined the traffic attribution and backdoor management practices used in phishkits. Our analysis shows that most of the practices used by phishkits to fingerprint users or block traffic are well known and have been used widely. We noticed excessive code reuse among phishkits in how their core functionalities operate. We conclude that most of the observed phishkits manifest similar behavior, which makes them easier to detect. We provide perspectives on how the security community might respond to such threats going forward.

\section*{Acknowledgment}
This project was supported by Microsoft AI Security, US National Science Foundation (NSF) under Grant No. 2219920, and Intergovernmental Personnel Act Independent Research \& Development Program. Opinions, findings, conclusions, or recommendations expressed in this material are those of the author(s) and do not necessarily reflect the views of the funding agencies.

\bibliographystyle{IEEEtran}
\bibliography{references,asiaccs}

\end{document}